\documentclass[sigconf,nonacm]{acmart}

\usepackage{microtype}
\usepackage{graphicx}
\usepackage{subfigure}
\usepackage{booktabs} 
\usepackage{amsmath}
\usepackage{xspace}
\usepackage{multirow}
\usepackage{xspace}
\usepackage{xcolor}
\usepackage{hyperref}
\usepackage{algorithm}
\usepackage{algorithmic}
\AtBeginDocument{%
  }

\newcommand{\oursys}{Rose\xspace}

\makeatletter
\renewcommand{\@ACM@checkaffil}{%
  \if@ACM@instpresent\else
    \ClassWarningNoLine{\@classname}
      {No institution present for an affiliation}%
  \fi
}
\makeatother

\begin{document}

\title{\oursys: Efficient Embodied RL through Graph Optimization and GPU Compute Resource Control}

\author{Wen Zeng}
\affiliation{\institution{Tsinghua University, Infinigence AI}}
\email{zengw25@mails.tsinghua.edu.cn}

\author{Ke Hong}
\affiliation{\institution{Tsinghua University}}

\author{Xu Fu}
\affiliation{\institution{Infinigence AI}}

\author{Zixiao Huang}
\affiliation{\institution{Tsinghua University}}

\author{Zehao Wang}
\affiliation{\institution{Tsinghua University}}

\author{Zhuoran Zhu}
\affiliation{\institution{Tsinghua University}}

\author{Yuanqing Wang}
\affiliation{\institution{Infinigence AI}}

\author{Bo Dai}
\affiliation{\institution{Infinigence AI}}

\author{Zhen Guo}
\affiliation{\institution{Infinigence AI}}

\author{Hao Lin}
\affiliation{\institution{Tsinghua University}}

\author{Chunyang Zhu}
\affiliation{\institution{Infinigence AI}}

\author{Boxun Li}
\affiliation{\institution{Infinigence AI}}

\author{Chao Yu}
\affiliation{\institution{Tsinghua University}}

\author{Quanlu Zhang}
\affiliation{\institution{Infinigence AI}}

\author{Guohao Dai}
\affiliation{\institution{Shanghai Jiaotong University}}

\author{Zhenhua Zhu}
\affiliation{\institution{Tsinghua University}}

\author{Yu Wang}
\affiliation{\institution{Tsinghua University}}
\correspondingauthor

\renewcommand{\shortauthors}{Zeng et al.}

\begin{abstract}
Embodied reinforcement learning (RL) coordinates policy inference, simulation, and model training with highly heterogeneous resource demands.
Although individual components frequently underutilize GPU compute during execution, existing systems only overlap work at coarse component boundaries (e.g., pipelining inference and simulation across simulator groups), leaving substantial low-activity GPU windows within subcomponent phases and nested loops.
Exploiting these fine-grained windows manually requires tedious dependency analysis and workload-specific coordination.
Furthermore, even when independent work is scheduled concurrently, cross-process GPU time slicing and streaming multiprocessor (SM) contention can negate the benefits of overlap.

We present \oursys, a runtime-aware system that automates fine-grained overlap and GPU resource management for embodied RL while preserving synchronous training semantics.
Through a lightweight programmable interface
that exposes execution hierarchy and state accesses, \oursys tracks runtime data readiness and GPU activity to perform a two-stage optimization.
First, a window-guided graph optimizer identifies low-activity windows and applies graph transformations to overlap complementary work across components and loop levels. 
Second, leveraging NVIDIA Multi-Process Service (MPS) for cross-process GPU concurrency, a slack-aware resource manager tunes process-level SM quotas by exploiting branch synchronization slack to mitigate interference on the critical path.
Both stages combine local profiling with end-to-end iteration validation to guarantee throughput gains with minimal tuning overhead.
Across diverse policies, simulators, RL algorithms, and GPU platforms, \oursys achieves 1.20--1.92$\times$ training throughput improvement and increases average SM activity by 23.8--39.3\% compared with the state-of-the-art systems.
\end{abstract}

\maketitle

\section{Introduction}
\label{sec:introduction}

Reinforcement learning is increasingly used
to train embodied policies for complex robotic
tasks.
Through trial-and-error interaction with simulators,
RL can improve task performance and execution
robustness~\cite{simplevlarl,WCRB,flare}.

\begin{figure}[t]
    \centering
    \includegraphics[width=\linewidth]{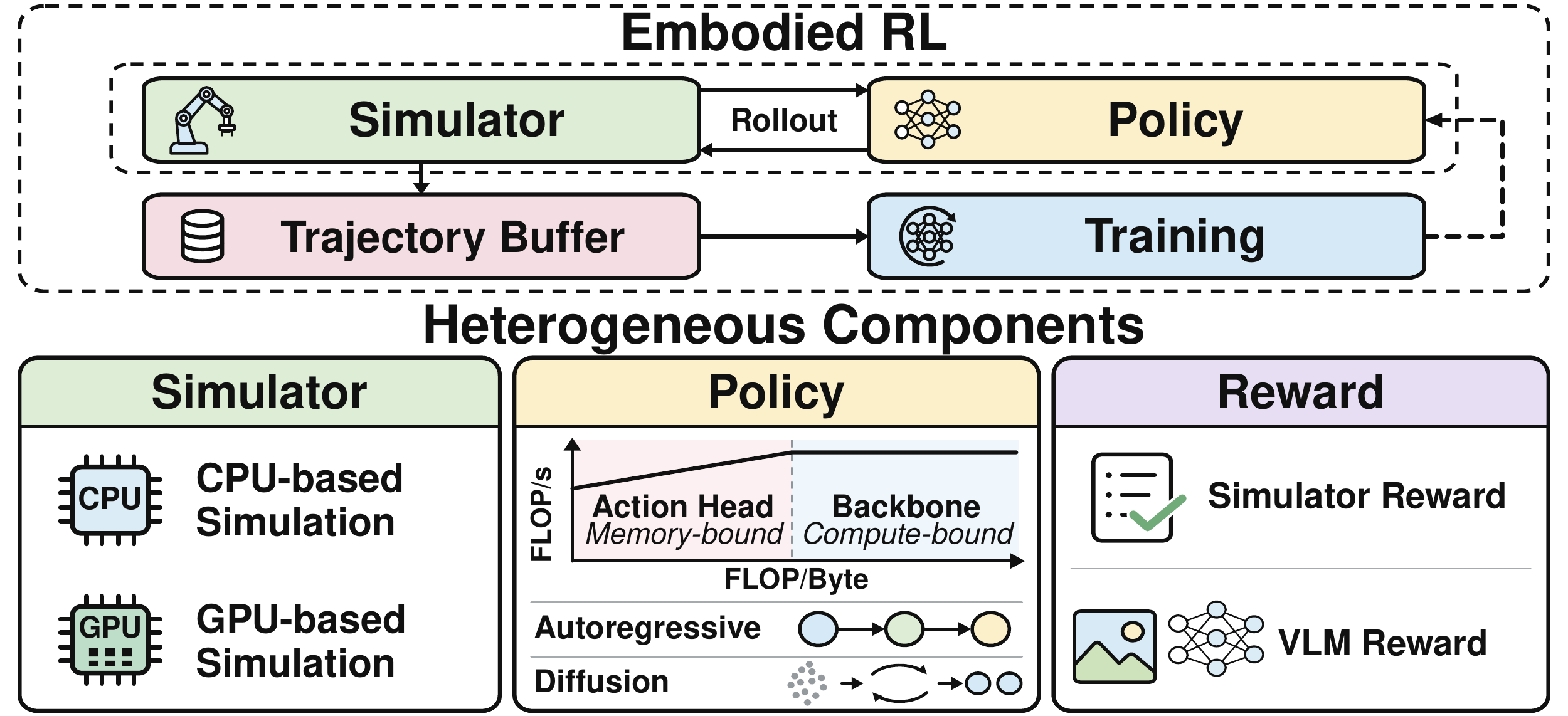}
    \vspace{-10pt}
    \caption{Embodied RL system and heterogeneous components.}
    \vspace{-10pt}
    \label{fig:RL-sys-diverse}
\end{figure}

Training these policies with RL requires coordinating multiple components with heterogeneous resource demands across rollout and policy training.
As illustrated in Figure~\ref{fig:RL-sys-diverse}, rollout combines policy inference, simulation, and reward computation, while policy training performs model updates.
Policy training and backbone inference require substantial GPU computation, while simulation combines physics computation and rendering with CPU and GPU demands that vary across simulators.
Moreover, a component's execution can be divided into \emph{phases} corresponding to its subcomponents.
GPU demand varies considerably within these phases, which leave GPU resources underused during parts of their execution.

This variety motivates placing components with complementary resource demands on the same GPUs to share compute resources, a placement strategy known as colocation.
Existing systems such as RLinf~\cite{rlinf} support flexible component placement and pipeline policy inference with simulation across simulator groups to improve utilization for colocation~\cite{rlux-vla}.
Nevertheless, profiling RLinf training a $\pi_{0.5}$~\cite{pi05} policy on LIBERO~\cite{libero} with this pipeline reveals substantial remaining windows.
Device-wide SM activity stays below 50\% for 38.8\% of the policy-simulator interaction loop execution time and 59.7\% of the complete iteration time (Figure~\ref{fig:SM-activity}).
Nsight Systems traces locate the interaction-loop windows in the action-head phase at the end of policy inference.

These pipelines schedule entire component invocations, leaving internal phases outside the scheduler's visibility.
Our key insight is that \emph{an independent phase that can exploit a low-activity window may belong to the same component, another component, or a different loop level}.
Here, independence means the phase can be moved into the target window without breaking data and state dependencies.
For example, the backbone phase of one pipeline group may overlap with the action-head phase of another group once the former's observations are ready.
Exploiting such opportunities requires exposing phase boundaries, intermediate-data readiness, and state constraints.

However, scheduling independent phases does not by itself ensure concurrent GPU execution.
Kernels submitted by separate component processes may still execute through GPU time slicing.
NVIDIA Multi-Process Service (MPS) enables concurrent kernel execution across processes~\cite{nvidia_mps}, but resource contention can offset the benefits of overlap.
The resulting slowdowns can affect downstream execution differently.
An earlier-finishing branch may have \emph{scheduling slack}: it can tolerate some slowdown without delaying dependent downstream phases.
This slack can guide which process to constrain to reduce interference on the branch that determines downstream progress.
Because quotas apply to entire processes, their effects extend beyond individual overlap regions and must be evaluated over the complete iteration.

We present \oursys, a runtime-aware system for optimizing execution graphs and GPU resource sharing in colocated embodied RL.
Through a programmable interface, developers wrap their existing implementations to expose execution regions and loops and annotate state accesses.
\oursys records invocation-level dependencies, intermediate-data readiness, and annotated state accesses at runtime.
The execution hierarchy provides the structure for locating transformation opportunities, while runtime records supply the constraints for checking their legality.
\oursys uses this information to construct overlap schedules without requiring developers to implement workload-specific scheduling or coordination.

\oursys performs optimization in two stages.
First, it identifies low-activity windows from GPU activity measurements and uses the execution hierarchy  to locate candidate phases and enclosing loops where graph transformations may expose overlap.
Generic transformations partition and reschedule the exposed regions while preserving data and state dependencies and training update boundaries.
Local profiling checks whether candidates shorten the completion time of affected regions.
Promising transformations are composed into candidate graphs.
Second, with the selected graph fixed, measured concurrent execution costs and branch slack guide process-level SM quota tuning.
Both stages retain changes only when complete-iteration measurements confirm an improvement.
Candidate limits and measurement reuse reduce tuning cost.
The evaluated policies use fixed inference steps per action chunk without variable-length autoregressive decoding.
This regularity supports keeping the selected graph and quotas fixed during subsequent training.

We evaluate \oursys across models, simulators, algorithms, and GPU platforms.
Against the strongest evaluated RLinf configuration and the asynchronous baseline RL-VLA$^3$, \oursys achieves 1.20--1.92$\times$ higher end-to-end throughput, measured in simulator steps/s over complete training iterations.
It also increases SM activity, averaged over time and across GPUs, by 23.8--39.3\%.
Our contributions are:
\begin{itemize}
    \item \textbf{A hierarchical execution model with
composable transformations.}
    We combine developer-exposed execution structure and state annotations with invocation-level dependencies and data readiness to support graph transformations that preserve dependencies and training update boundaries.

    \item \textbf{Window-guided graph optimization.}
    We use low-activity windows and execution hierarchy to discover overlap opportunities, check candidate transformations through local measurements, and select execution graphs through complete-iteration evaluation.

    \item \textbf{Slack-aware GPU resource optimization.}
    We combine branch slack and measured concurrent execution costs to guide process-level SM quota tuning, using a coarse-to-fine search and accounting for quota effects beyond the targeted overlap region.
\end{itemize}
\section{Background and Related Works}
\label{sec:background}

\subsection{Embodied Reinforcement Learning}
Embodied RL combines three main components:
(1) \emph{policy inference}, (2) \emph{environment simulation}, and (3) \emph{policy optimization}, denoted as Policy, Simulation, and Training in Figure~\ref{fig:RLinf-sys-pipe}, respectively. 
The first two components jointly perform
rollout collection, \textit{i.e.}, 
the policy generates actions, while the
environment executes them and returns new
observations and rewards. 
Repeated interactions produce trajectories, which
policy optimization uses to update the policy toward
maximizing rewards.
Colocated placement shares GPUs among simulator, policy inference, and training.
Disaggregated placement assigns components to separate GPU subsets.
Hybrid placement assigns simulation and inference to separate GPU subsets during rollout and reuses GPU resources for training.
Simulator-group pipelining interleaves inference and simulation across simulator groups.
In an asynchronous configuration, components execute asynchronously on separate GPU subsets, allowing
rollout and training to overlap.


\subsection{Policy Inference.}
Policy inference combines a vision-language backbone with
an action generation module to map observations and task
instructions to actions~\cite{pi0,pi05,groot,openvla,openvla_oft}.
Two stages have different GPU demands, and action generation
may contain repeated decoding or refinement steps.
Fixed output sizes and generation budgets make the execution
structure tend to be regular.

\subsection{Environment Simulation.}
Environment simulation combines physics stepping, rendering,
reward computation, and episode resets.
LIBERO and RoboCasa use CPU physics with GPU rendering
~\cite{libero,robocasa,robosuite}, while ManiSkill3 and
BEHAVIOR-1K support GPU physics and rendering
~\cite{maniskill3,behavior1k}.
Environment instances run in parallel, but interaction steps
within each instance retain sequential data and state dependencies.

\subsection{Policy Optimization.}
Policy optimization performs repeated minibatch updates over
collected trajectories~\cite{ppo,deepseekmath,simplevlarl},
involving GPU-intensive forward and backward computation.
Trajectory availability and reuse, policy-version requirements,
and parameter-update order constrain its overlap with rollout.

\subsection{Related Works}

\begin{figure}[t]
    \centering
    \includegraphics[width=\linewidth]{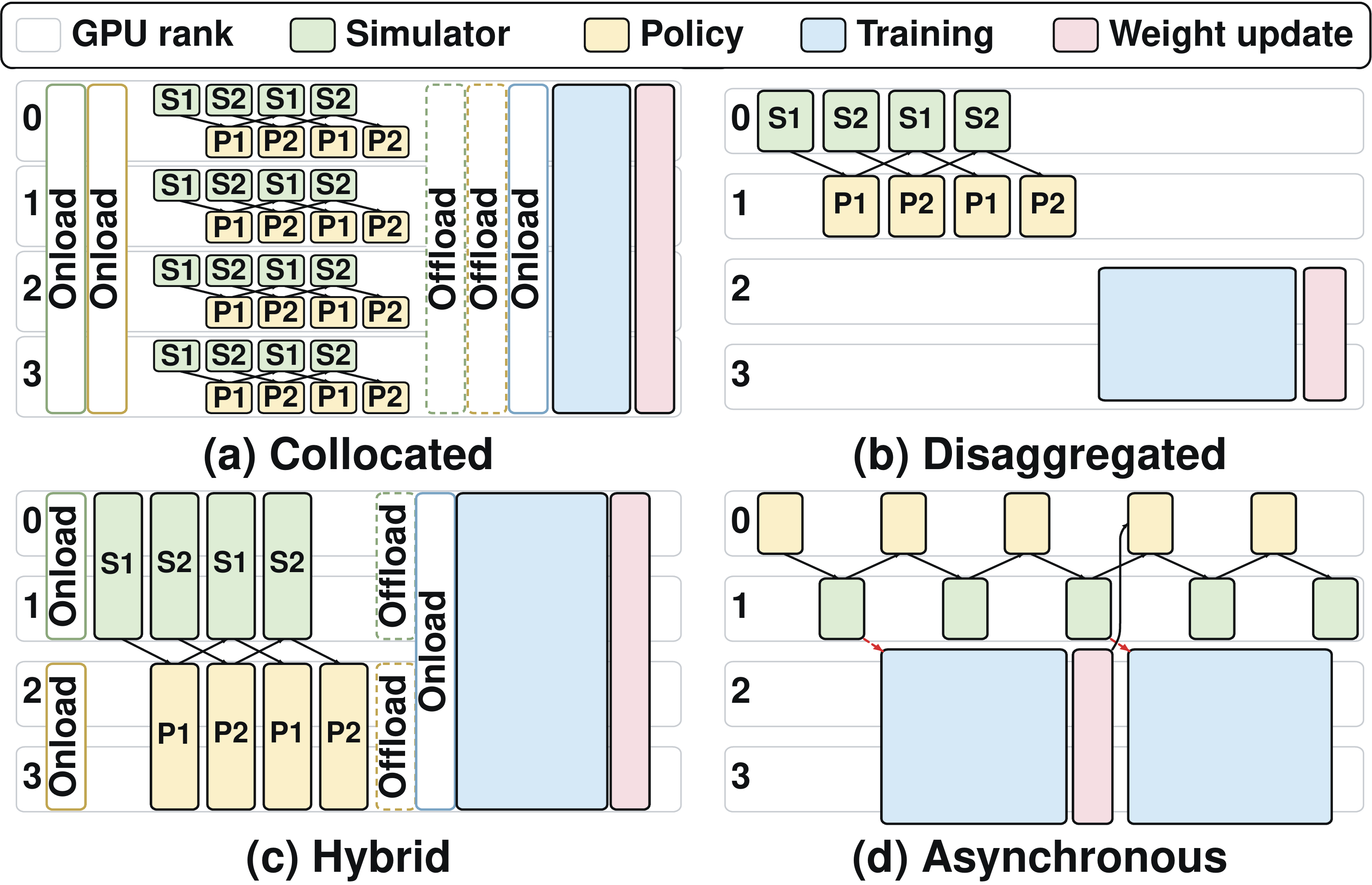}
    \vspace{-15pt}
    \caption{Embodied RL execution modes. }
    \vspace{-10pt}
    \label{fig:RLinf-sys-pipe}
\end{figure}

\textbf{RL Systems.}
OpenRLHF~\cite{openrlhf}, HybridFlow~\cite{hybridflow}, and ReaL~\cite{real}
improve RL efficiency through flexible placement and parameter resharding
between rollout and training, while RLHFuse~\cite{rlhfuse} and
AReaL~\cite{areal} overlap stages via stage fusion and asynchronous rollout,
respectively. RLinf~\cite{rlinf} extends such flexibility to embodied RL
(Figure~\ref{fig:RLinf-sys-pipe}).
These systems choose among manually designed execution modes, whereas \oursys
derives overlap from the workflow graph.

\noindent\textbf{Asynchronous Embodied RL Systems.}
RL-VLA$^3$~\cite{RL-VLA3} and EBRL~\cite{EBRL} modify RL algorithms for
asynchronous execution, overlapping simulation, inference, and training,
with EBRL further adjusting CPU and GPU quotas at runtime.
Such modifications potentially compromise convergence.
In contrast, \oursys preserves synchronous algorithms and derives overlap
via graph transformations, without changing the
algorithm's semantics.

\noindent\textbf{Graph-based Execution and Scheduling.}
Dataflow systems~\cite{dryad,naiad,tensorflow} express computation as
dependency graphs, and optimizers~\cite{taso,unity,alpa} and
schedulers~\cite{heft,rammer} search for semantics-preserving transformations and
map task DAGs onto heterogeneous resources.
\oursys applies these ideas to embodied RL, where nested loops and policy-version
dependencies constrain valid transformations.

\begin{figure}[t]
    \centering
    \includegraphics[width=\linewidth]{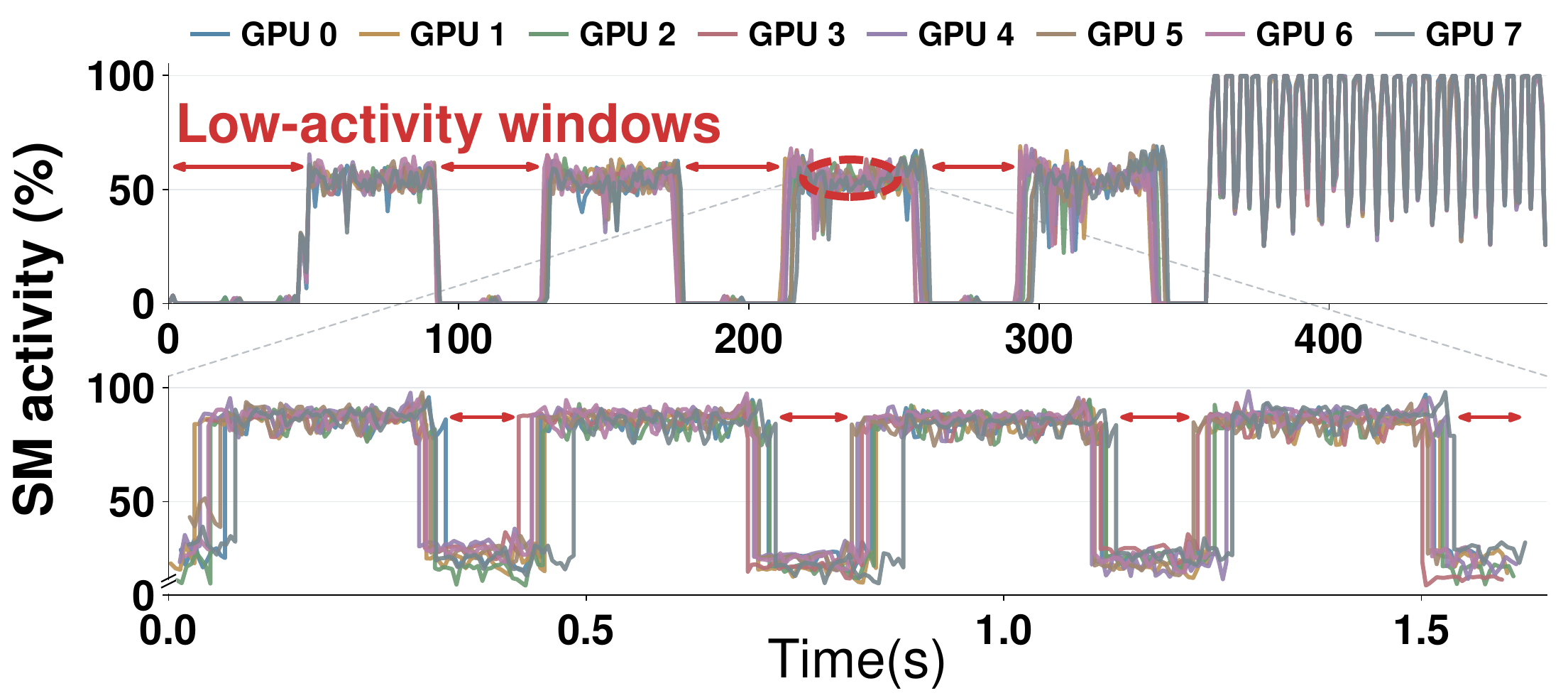}
    \vspace{-20pt}
    \caption{SM activity over a complete training iteration with simulator-group pipeline. The lower panel shows policy-simulator interaction spread from the upper rollout.}
    \vspace{-10pt}
    \label{fig:SM-activity}
\end{figure}

\section{Motivation}
\label{sec:motivation}


\subsection{Low-Activity Windows Across Execution Levels}



Whole-component scheduling leaves GPU resources underused within components and across loop levels, as illustrated in Figure~\ref{fig:SM-activity}.
During one group's action-head execution, another group's backbone inputs may already be ready.
Across rollout-epoch boundaries, training may overlap with environment reset when trajectory data are available and state accesses are independent.
Exploiting these fine-grained overlap opportunities raises two key challenges:

\noindent \textbf{Challenge-1: Determining legal overlap plans.}
Reordering must preserve data and state dependencies, the policy versions used for rollout, and training update boundaries.


\noindent \textbf{Challenge-2: Selecting beneficial overlap plans.}
Partitioning and rescheduling introduce multiple candidate graphs.
Downstream readiness can affect the benefits of combined transformations, and exhaustive evaluation is costly.

\subsection{GPU Concurrency and Resource Interference}



\begin{figure}[t]
    \centering
    \includegraphics[width=\linewidth]{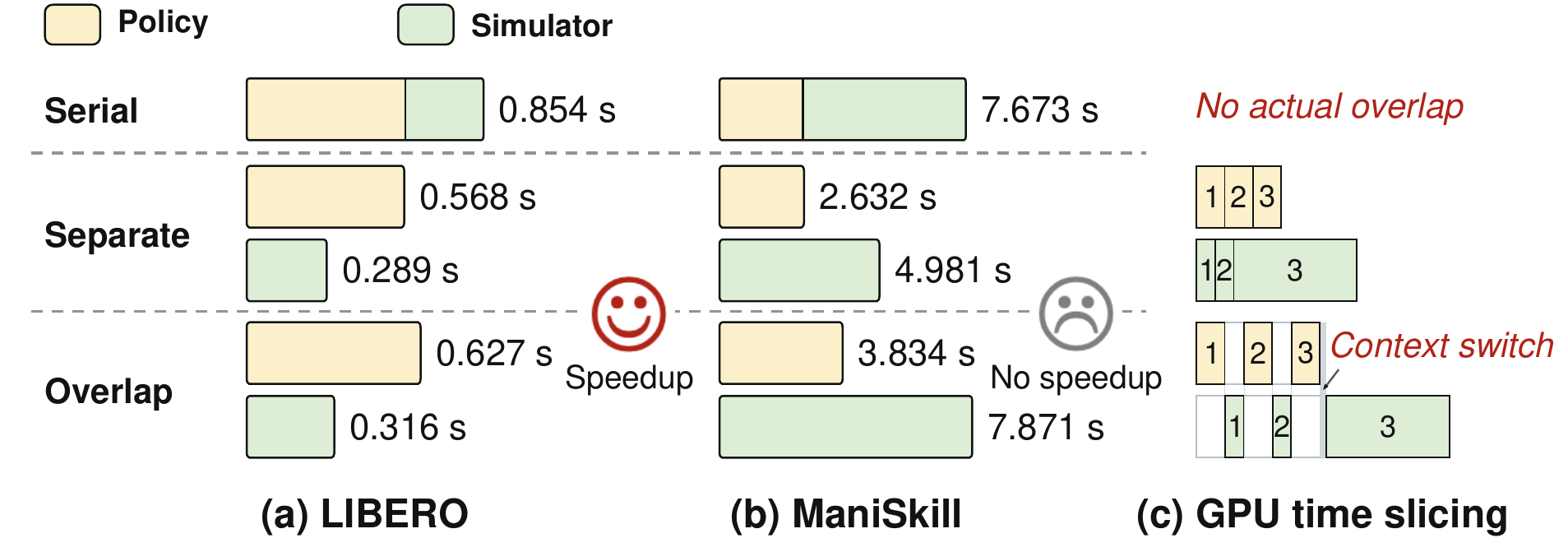}
    \vspace{-25pt}
    \caption{Different execution modes and latency of CPU-based and GPU-based simulators. (a) $\pi_{0.5}$ and LIBERO, batch=16. (b) $\pi_{0.5}$ and ManiSkill, batch=80. (c) Schematic illustration of time-sliced GPU execution across CUDA contexts without MPS.}
    \vspace{-10pt}
    \label{fig:time-slice}
\end{figure}

\noindent \textbf{GPU time slicing.}
GPU execution may remain time-sliced even when processes overlap.
Figure~\ref{fig:time-slice} compares serial and overlapping execution of policy inference and simulation.
Without MPS, overlap reduces their joint completion time for LIBERO, but provides no reduction for ManiSkill.
For ManiSkill, we observe time-sliced GPU execution across separate CUDA contexts despite process-level overlap, as illustrated in Figure~\ref{fig:time-slice}(c).

\begin{table*}[t]
    \centering
    \small
    \vspace{-10pt}
    \caption{Policy inference and simulator execution times.
    The reported speedup is
    $T_{\mathrm{serial}}/
    \max(T_{\mathrm{inf,overlap}},T_{\mathrm{sim,overlap}})$.}
    \label{tab:overlap}
    \small
    \begin{tabular}{llcccccc}
        \toprule
        Simulator &
        MPS &
        Inf. alone (s) &
        Sim. alone (s) &
        Serial (s) &
        Inf. overlap (s) &
        Sim. overlap (s) &
        Speedup \\
        \midrule
        LIBERO &
        Off &
        0.568 & 0.289 & 0.854 &
        0.627 & 0.316 & 1.36$\times$ \\
        ManiSkill &
        On &
        2.632 & 4.981 & 7.673 &
        2.661 & 7.192 & 1.07$\times$ \\
        \bottomrule
    \end{tabular}
    \vspace{-10pt}
\end{table*}


\noindent \textbf{Interference under GPU concurrency.}
NVIDIA MPS enables cross-process GPU concurrency~\cite{nvidia_mps}, but colocated workloads still share execution resources.
Table~\ref{tab:overlap} shows that, with MPS enabled, overlapping policy inference and ManiSkill simulation increases their execution time by 1.1\% and 44.4\%, respectively, relative to isolated execution.
As simulation determines their joint completion time, the resulting speedup is limited to 1.07$\times$.
Inference finishes earlier than simulation, leaving  branch slack before joint completion.
To reduce interference raises the following challenge:

\noindent \textbf{Challenge-3: Tuning SM quotas under interference.}
Quota tuning must balance branch slack, concurrent execution costs, and effects beyond the targeted overlap region.

\section{Design}
\label{sec:design}


\subsection{Overview}
\label{sec:design-overview}

\begin{figure}[t]
    \centering
    \includegraphics[width=\linewidth]{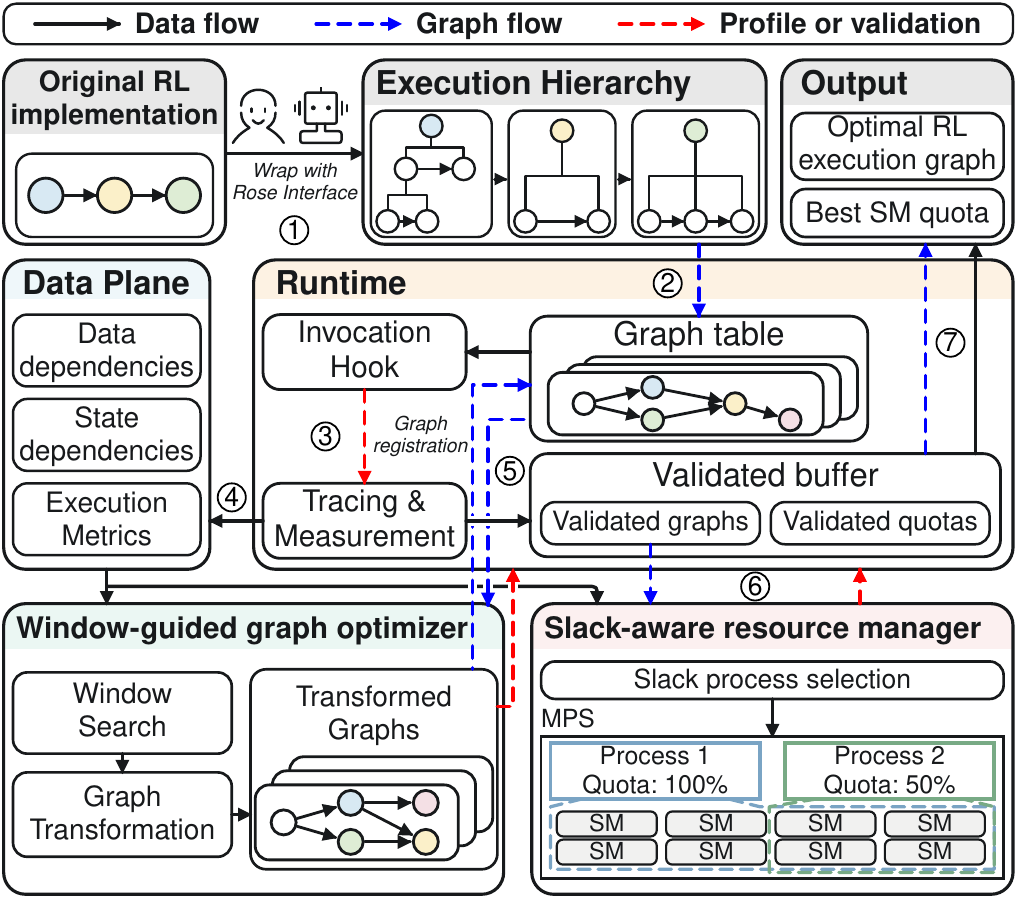}
    \vspace{-20pt}
    \caption{Overview of \oursys.
    \textcircled{\scriptsize 1}
    Code wrapping to expose hierarchy.
    \textcircled{\scriptsize 2}
    The original graph registration.
    \textcircled{\scriptsize 3}
    Graph execution dispatch and measurement.
    \textcircled{\scriptsize 4}
    Collects dependencies and metrics for optimization.
    \textcircled{\scriptsize 5}
    The graph optimizer retrieves graphs, transforms them, and registers back.
    Validated graphs are retained and this may run for multiple rounds.
    \textcircled{\scriptsize 6}
    The resource manager retrieves the best graph, tuning the selected process SM quota, and stores the best quota back.
    \textcircled{\scriptsize 7}
    The buffer outputs final plan for subsequent RL training.
    }    
    \vspace{-10pt}
    \label{fig:overview}
\end{figure}

As illustrated in Figure~\ref{fig:overview}, \oursys exposes execution hierarchy and forms execution graphs at runtime (\S\ref{sec:execution-model and data-plane}). With the metrics of the execution graph, \oursys performs a two-stage optimization:\emph{graph-based overlap} and \emph{slack-aware resource management}.
The graph optimizer identifies low-activity nodes, searches for complementary work, and transforms the execution graphs to enable overlap (\S\ref{sec:Graph-based optimization}).
With the execution graph fixed, the resource manager tunes the quota of a GPU process with slack to reduce interference between overlapping processes (\S\ref{sec:resource-optimization}).

\begin{figure}[t]
    \centering
    \includegraphics[width=\linewidth]{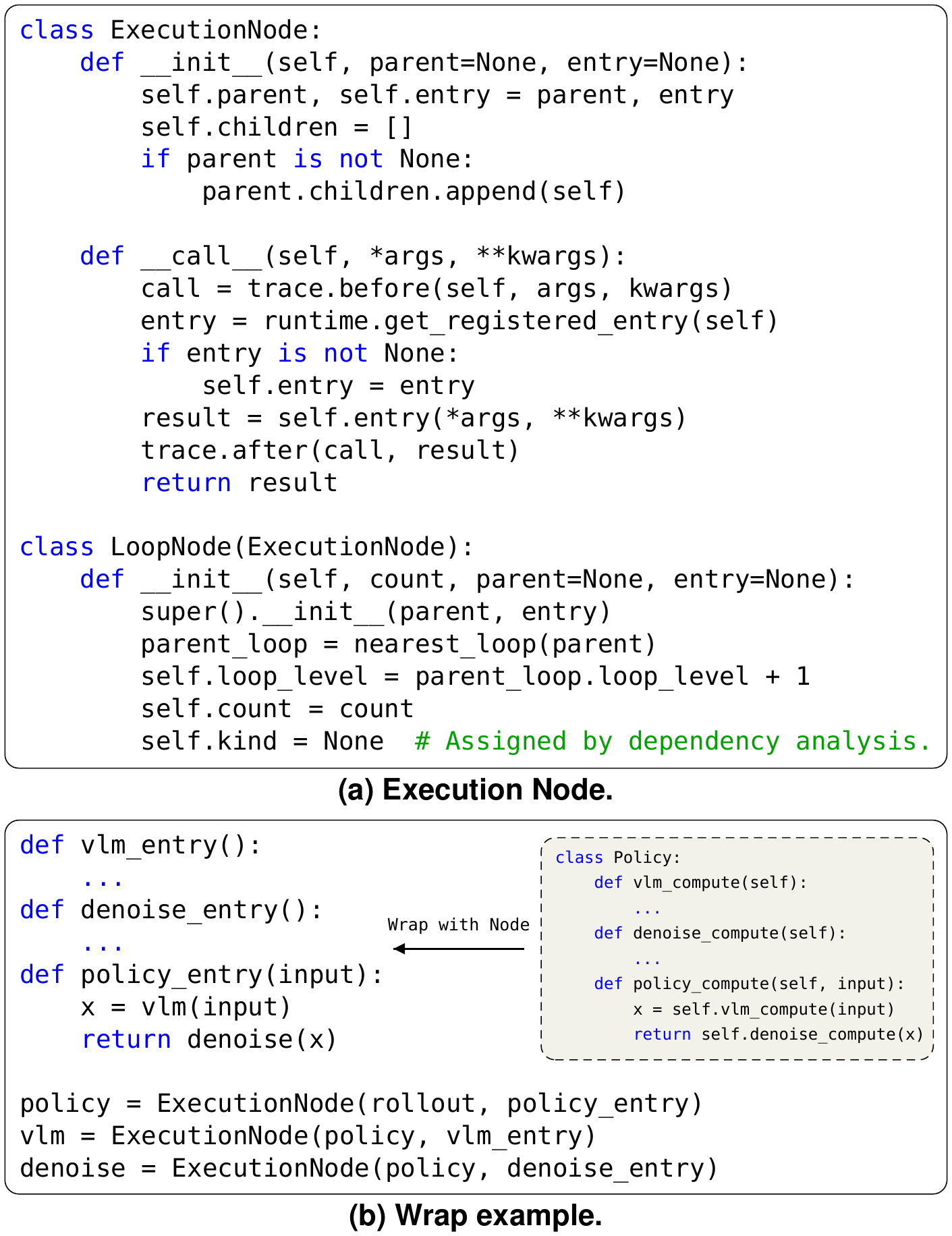}
    \vspace{-25pt}
    \caption{(a) Execution node wrapper. (b) An example of wrapping policy code, can also wrap new component code.}
    \vspace{-15pt}
    \label{fig:Execution-Nodes}
\end{figure}


\begin{figure}[t]
    \centering
    \includegraphics[width=\linewidth]{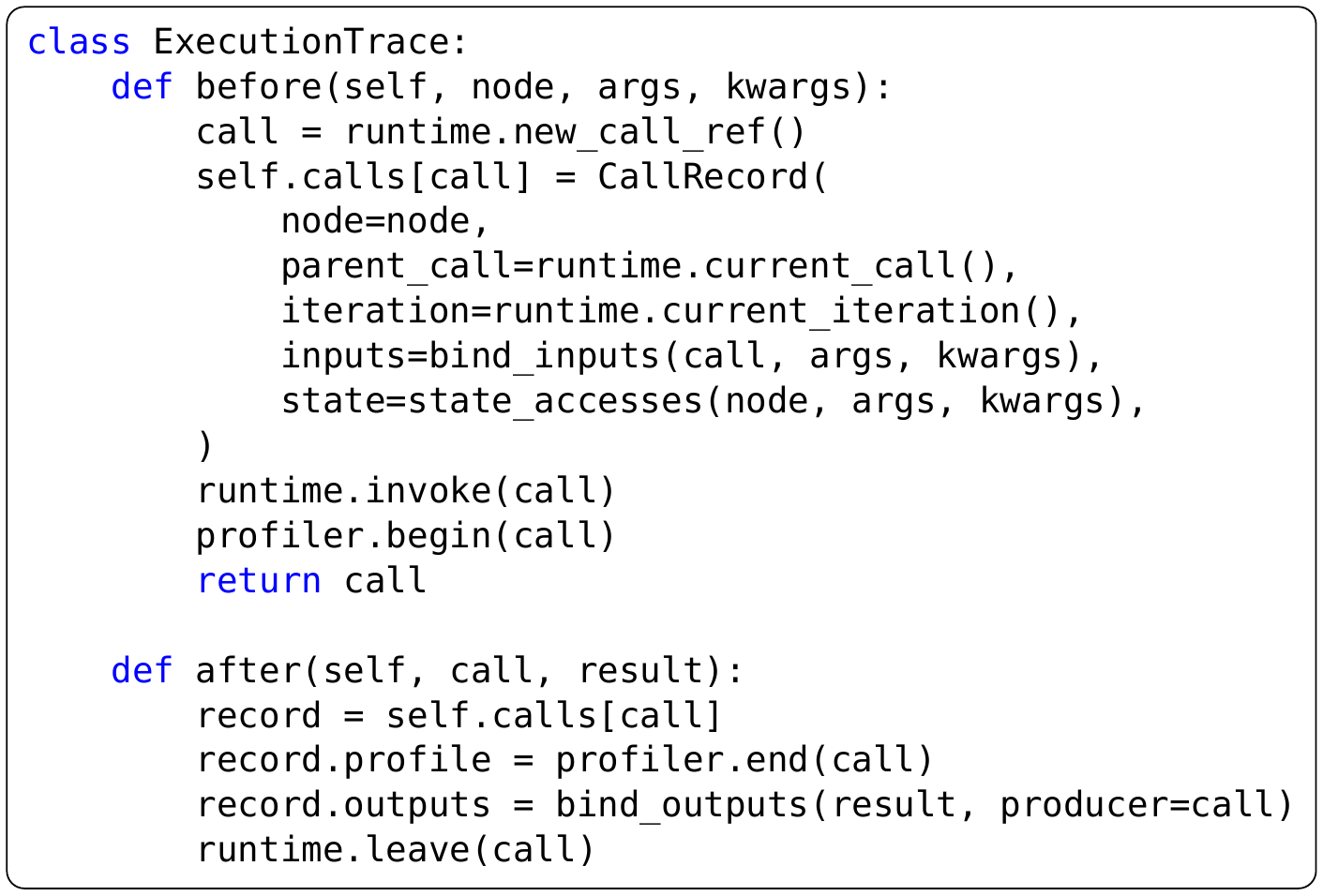}
    \vspace{-20pt}
    \caption{Execution trace wrapper for invocation recording.}
    \vspace{-10pt}
    \label{fig:Execution-Trace}
\end{figure}

\subsection{Programmable Execution Model}
\label{sec:execution-model and data-plane}

\oursys provides a hierarchical execution graph representation of existing RL programs through a programming interface.
Graph construction combines the exposed structure with state annotations and invocation records to establish dependencies and loop independence, enabling overlap across components and execution levels.

\subsubsection{Hierarchical Execution Graph Representation}


\oursys provides \texttt{ExecutionNode} and \texttt{LoopNode} to represent execution as hierarchical execution graphs, as shown in Figure~\ref{fig:Execution-Nodes}(a).
An \texttt{ExecutionNode} wraps a code block in a component worker with a callable entry and may expose an internal graph of child nodes, data bindings, and execution constraints.
Figure~\ref{fig:Execution-Nodes}(b) illustrates this wrapping for policy code.
Parent-child relations form an execution hierarchy rooted at the RL iteration loop, supporting analysis at different granularities.
\texttt{LoopNode} extends \texttt{ExecutionNode}:
its internal graph describes the loop body, while its iteration count specifies repetition.
Loops with conditional termination, such as \texttt{while} loops, are wrapped with \texttt{ExecutionNode} to retain the native loop control.
Accordingly, edges represent execution constraints that preserve required data and state dependencies.

An invocation is a distinct execution of a node, which recurs across loop iterations or input groups.
A node defines the exposed computation and callable
entries, while invocation records associate each execution with
its parent invocation, loop iteration, data references, and
state accesses.

\subsubsection{Execution Graph Construction}
\oursys constructs execution graphs by combining the exposed
hierarchy with invocation-level bindings.

\noindent \textbf{Invocation recording.}
Figure~\ref{fig:Execution-Trace} shows invocation recording through \texttt{ExecutionTrace}, using the hooks called in Figure~\ref{fig:Execution-Nodes}(a).
Each record identifies the structural node, parent invocation, and current iteration.
The \texttt{before} hook binds inputs to data references and records state accesses, while \texttt{after} binds outputs.
These records link invocations to the hierarchy and state dependency analysis.

\noindent \textbf{Data dependencies.}
\oursys uses data references to identify which invocation produces
each input, establishing producer-consumer dependencies. For loops, it records outputs at the end of each
iteration, allowing consumers to proceed once their required
data are ready without waiting for the entire loop to finish.
For example, trajectory data from each rollout epoch can become
available before the rollout returns the full collection.

\noindent \textbf{State dependencies.}
Beyond explicit inputs and outputs, nodes may access shared objects such as model weights and simulator state.
\oursys provides three annotations to describe such accesses:
\texttt{read} existing state, \texttt{update} new state from its previous value, and \texttt{refresh} new state without inheriting its previous value.
Each annotation applies to a specific state object. The system binds these annotations to invocation records to derive state dependencies.

\noindent \textbf{Loop independence.}
\oursys analyzes data and state dependencies across iterations to determine loop independence.
For a \texttt{LoopNode}, the classification is assigned by dependency analysis, as indicated by its \texttt{kind} field in Figure~\ref{fig:Execution-Nodes}.

A simulator reset annotated as \texttt{refresh} indicates that the new epoch's episode state does not inherit the previous epoch's state value.
If the remaining data and state constraints also permit independent execution, the system identifies the loop as a \emph{free loop}.
A loop with dependencies across iterations is classified as a \emph{dependent loop}.
Loops with conditional termination, such as \texttt{while} loops, are classified as dependent loops without an iteration count.

The resulting execution graphs and loop independence information provide the graph optimizer with the execution structure and constraints for reorganization.

\begin{figure}[t]
    \centering
    \includegraphics[width=\linewidth]{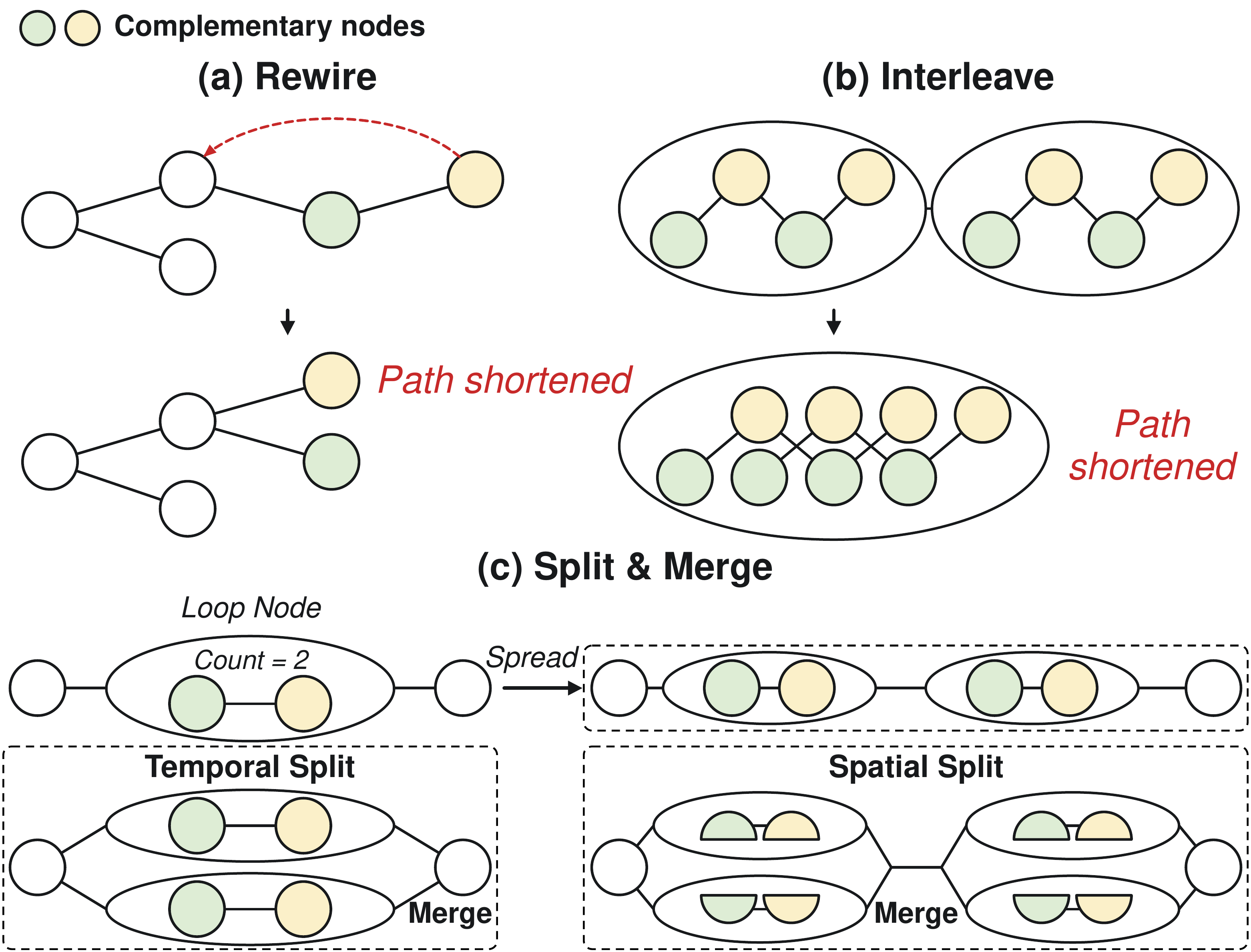}
    \vspace{-20pt}
    \caption{Transformation APIs.}
    \vspace{-10pt}
    \label{fig:APIs}
\end{figure}

\begin{figure*}[t]
    \centering
    \includegraphics[width=\linewidth]{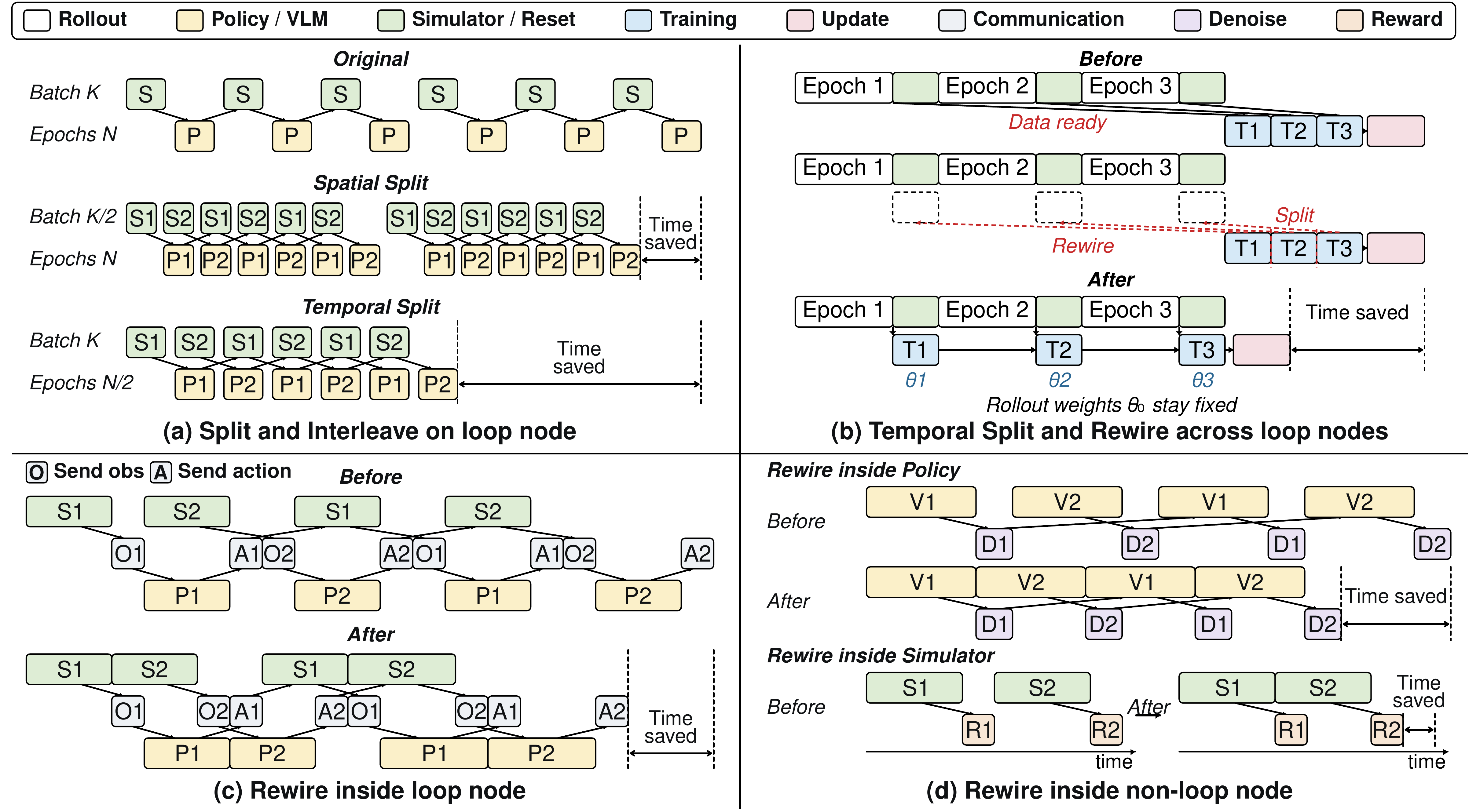}
    \vspace{-20pt}
    \caption{Optimization examples. Black arrows indicate execution dependencies. (a) Split simulator-policy interact loop and interleave them. (b) Split training batch loop and rewire it to overlap simulator reset. (c) Rewire to overlap communication inside simulator-policy interaction node. (d) Rewire inside a component node between independent component graphs. Reward computation is a part of complete simulator computation.}
    \vspace{-10pt}
    \label{fig:transform_examples}
\end{figure*}

\subsection{Window-guided Graph Optimization}
\label{sec:Graph-based optimization}

Building on the execution graphs, \oursys identifies low-activity windows and searches for work that can execute during these windows while respecting data and state dependencies.
It constructs overlap schedules through composable graph transformations (Figure~\ref{fig:APIs}), uses local measurements to guide candidate selection and composition, and retains only changes validated to reduce end-to-end iteration time.
Graph optimization examples are illustrated in Figure~\ref{fig:transform_examples}.

\subsubsection{Graph Transformation APIs}
\label{sec:transformations}
\oursys provides four execution graph transformation APIs:
Rewire, Split, Merge, and Interleave.
The transformations reorganize execution graphs while reusing nodes' callable entries and preserving required data and state dependencies.

\noindent \textbf{Rewire.}
Rewire changes the execution order of existing nodes while preserving required data and state dependencies.
It applies to both loop and non-loop nodes whose inputs and state permit the new placement. 



\noindent \textbf{Split.}
Split constructs separately schedulable executions of a node or its internal graph, each bound to a data partition or an independent group of loop iterations.
\textbf{Spatial split} partitions batched computation into groups with separable data and state while preserving each group's iteration order.
\textbf{Temporal split} organizes independent iterations of a \emph{free loop} into separate execution graphs while preserving the original batch sizes.

\noindent \textbf{Merge.}
Merge collects results from the groups or execution graphs produced by Split to reconstruct the input required by the next unsplit node.

\noindent \textbf{Interleave.}
Interleave staggers nodes across the execution graphs produced by Split to enable overlap while preserving dependencies within each graph.
For example, a legal Split of a policy-simulator step produces data groups A and B, and Interleave makes simulator(A) overlap policy(B), while each group retains its policy-before-simulator ordering.

\subsubsection{Overlap Candidate Discovery}
\label{sec:candidate-discovery}

\oursys uses the execution intervals of node invocations as candidate overlap windows. To limit candidate trials, it first ranks nodes by duration and SM activity over these windows, then searches for work that can overlap under data and state constraints, and finally orders feasible candidates by duration matching.



\noindent \textbf{Ranking low-activity nodes.}
For a schedulable node $v$, $D_v$ is its total invocation duration within one training iteration, and $\bar{a}_v$ is the time-weighted
average SM activity observed over these invocations.
Motivated by NVIDIA's profiling metrics documentation~\cite{nvidia_dcgm_profiling}, \oursys excludes nodes with $\bar{a}_v \geq 0.8$ and prioritizes those with $\bar{a}_v < 0.5$ over those with $0.5 \leq \bar{a}_v < 0.8$.
Within each activity tier, \oursys ranks nodes by decreasing
$S(v)=D_v(1-\bar{a}_v)$.
The resulting rank $\pi(v)$ incorporates both the activity-tier
priority and the ordering within each tier. 
Eligible nodes must also satisfy $D_v \geq \min\{\tau_a,\tau_r T_{\mathrm{iter}}\}$, where $\tau_a$ is an absolute duration threshold and $\tau_r$ specifies a fraction of the observed iteration time $T_{\mathrm{iter}}$.




\noindent \textbf{Candidate search.}
For each \emph{target node} $v$ selected by this ranking, \oursys searches for a \emph{candidate node} $w$ whose work can overlap with $v$ through feasible graph transformations.
When $v$'s parent is a \texttt{LoopNode}, \oursys first searches its internal graph and tries Split followed by Interleave.
Temporal split uses independent iterations from the parent loop or legal free ancestors, considered from nearest to farthest.
If no free loop exists, \oursys uses Spatial split.
It then searches the loop's parent's internal graph in this order:
other movable loops, splittable downstream non-loop consumers, and unsplittable non-loop nodes that permit Rewire.

When $v$'s parent is non-loop, \oursys first checks whether $v$ can be repositioned, then searches other nodes in the parent's internal graph.
If no feasible candidate exists, it searches the parent's siblings.
Feasibility requires checking transformation applicability, data and state constraints, and estimated capacity.

\noindent \textbf{Candidate ordering.}
Candidate ordering uses node rank $\pi(v)$ followed by
increasing duration mismatch
$\delta(v,w)=|T(v)-\widehat{T}(w)|$.
Here, $T(v)$ is the mean invocation duration of $v$, representing
its average window duration, and $\widehat{T}(w)$ estimates
the candidate's duration after any required Split.
This ordering determines which alternatives receive measurements
under the per-node trial budget.

\subsubsection{Candidate Evaluation and Graph Composition}
\label{sec:runtime-graph-optimization}

\begin{algorithm}[t]
\caption{Window-guided graph optimization}
\label{alg:graph-optimization}
\small
\begin{algorithmic}[1]

\REQUIRE Feasible graph $G_0$;
         settings $\mathbf{M},\boldsymbol{\tau}$
\REQUIRE Target limit $K\geq1$;
         alternative limit per target $L\geq1$
\REQUIRE Round limit $R\geq1$;
         improvement threshold $\epsilon\in(0,1)$
\ENSURE Graph $G$ selected for subsequent training

\STATE $G\gets G_0$

\FOR{$r=1,\ldots,R$}

    \STATE $m_{\mathrm{ref}}\gets
        \mathrm{RunTraining}(G)$;
        $\mathcal{D}\gets[\,]$

    \FORALL{$v\in
        \mathrm{Top_K}
        (G,S(v),\boldsymbol{\tau})$}

        \STATE $\Phi_v\gets
            \mathrm{Top}_{L}
            \bigl(\mathrm{FeasibleTransform}(G,v);
                  (\pi,\delta)\bigr)$

        \STATE \textbf{if} $\Phi_v=\varnothing$
            \textbf{then continue}

        \STATE $A_v\gets
            \mathrm{CommonGraph}(G,v,\Phi_v)$

        \STATE $c_0\gets A_v$

        \STATE $\mathcal{C}_v\gets
            \{c_0\}\cup
            \{\mathrm{Transform}(A_v,\phi):
              \phi\in\Phi_v\}$

        \FORALL{$c\in\mathcal{C}_v$}

            \STATE $m_c\gets
                \mathrm{MeasureLocal}
                \bigl(c)$

        \ENDFOR

        \STATE $\mathcal{D}[v]\gets
            \bigl(c\mapsto m_c:c\in\mathcal{C}_v\bigr)$

    \ENDFOR

    \STATE $G_{\mathrm{comb}}\gets
        \mathrm{Combine}(G,\mathcal{D},\mathbf{M})$

    \IF{$G_{\mathrm{comb}}=G$ \OR
        $\neg\mathrm{Valid}(G_{\mathrm{comb}})$}
        \STATE \textbf{break}
    \ENDIF

    \STATE $m_{\mathrm{comb}}\gets
        \mathrm{RunOrReuse}(G_{\mathrm{comb}})$

    \IF{$m_{\mathrm{comb}}.\mathrm{time}\geq
        (1-\epsilon)m_{\mathrm{ref}}.\mathrm{time}$}
        \STATE \textbf{break}
    \ENDIF

    \STATE $G\gets G_{\mathrm{comb}}$

\ENDFOR

\STATE \textbf{return} $G$

\end{algorithmic}
\end{algorithm}

\begin{algorithm}[t]
\caption{Slack-aware GPU resource optimization}
\label{alg:resource-optimization}
\small
\begin{algorithmic}[1]

\REQUIRE Fixed graph $G$; target process $f$;
         comparison graph $A$
\REQUIRE Initial numerator $k_0$;
         quota size $Q$
\REQUIRE Coarse-step multiplier $\rho\geq2$
\ENSURE Selected quotas $\mathbf{q}^{*}$
        and measurement $m^{*}$

\STATE $k_b\gets k_0$
\STATE $\mathcal{H}\gets\{k_b\mapsto t(k_b)\}$

\item[] \COMMENT{Coarse search}

\WHILE{$k_b>1$}
    \STATE $k\gets\max(1,k_b-\rho)$;
           $\mathcal{H}[k]\gets t(k)$
    \STATE \textbf{if}
        $\mathcal{H}[k]>\mathcal{H}[k_b]$
        \textbf{then break}
    \STATE $k_b\gets k$
\ENDWHILE

\item[] \COMMENT{Fine search around the anchor $k_b$}

\FORALL{$d\in\{-1,+1\}$}
    \STATE $k\gets k_b+d$
    \WHILE{$1\leq k\leq Q$ \AND
           $k\notin\mathrm{dom}(\mathcal{H})$}
        \STATE $\mathcal{H}[k]\gets t(k)$
        \STATE \textbf{if}
            $\mathcal{H}[k]>\mathcal{H}[k_b]$
            \textbf{then break}
        \STATE $k\gets k+d$
    \ENDWHILE
\ENDFOR

\item[] \COMMENT{End-to-end validation}

\STATE $k^{*}\gets
    \underset{k\in\mathrm{dom}(\mathcal{H})}
    {\arg\min}\;\mathcal{H}[k]$
\STATE $(\mathbf{q}^{*},m^{*})\gets
    \mathrm{Validate}(k^{*})$
\STATE \textbf{return} $(\mathbf{q}^{*},m^{*})$

\end{algorithmic}
\end{algorithm}

Since local improvements do not guarantee reduced iteration time,
Algorithm~\ref{alg:graph-optimization} combines local measurements for graph transformation and complete-iteration validation to accept or reject the result.

All targets in a round use the same reference graph $G$, the
complete execution graph of an RL iteration.
For each target $v$, \texttt{FeasibleTransform}$(G,v)$ performs
the candidate search described above and returns feasible
transformation alternatives $\phi(v,w)$ for overlapping $v$
with $w$. The retained alternatives share a \emph{comparison graph}
$A_v$: the internal graph of the lowest common ancestor enclosing
all nodes affected by these alternatives.
\texttt{Transform}$(A_v,\phi)$ applies a retained alternative
to $A_v$, producing a candidate local graph $c$.
Both the transformed graphs and the unchanged graph $c_0=A_v$
are measured over this common comparison graph.
Their completion times account for downstream readiness,
reconstruction, and synchronization.

\texttt{Combine} uses the measured reductions in completion time
to greedily compose changes that can be applied together,
selecting at most one candidate per target.
Combined graphs are ranked by estimated cumulative benefit
and checked against the memory settings $M$ before complete-iteration
validation. A combined graph is accepted only when a valid
observation shows a relative reduction in iteration time exceeding
$\epsilon$ compared with $G$.
\texttt{RunOrReuse} reuses only end-to-end observations of the
same complete graph.
Each complete-iteration trial advances training and updates the
weights.
Its execution trace, time, and validity guide subsequent
optimization rounds.

\begin{figure}[t]
    \centering
    \includegraphics[width=\linewidth]{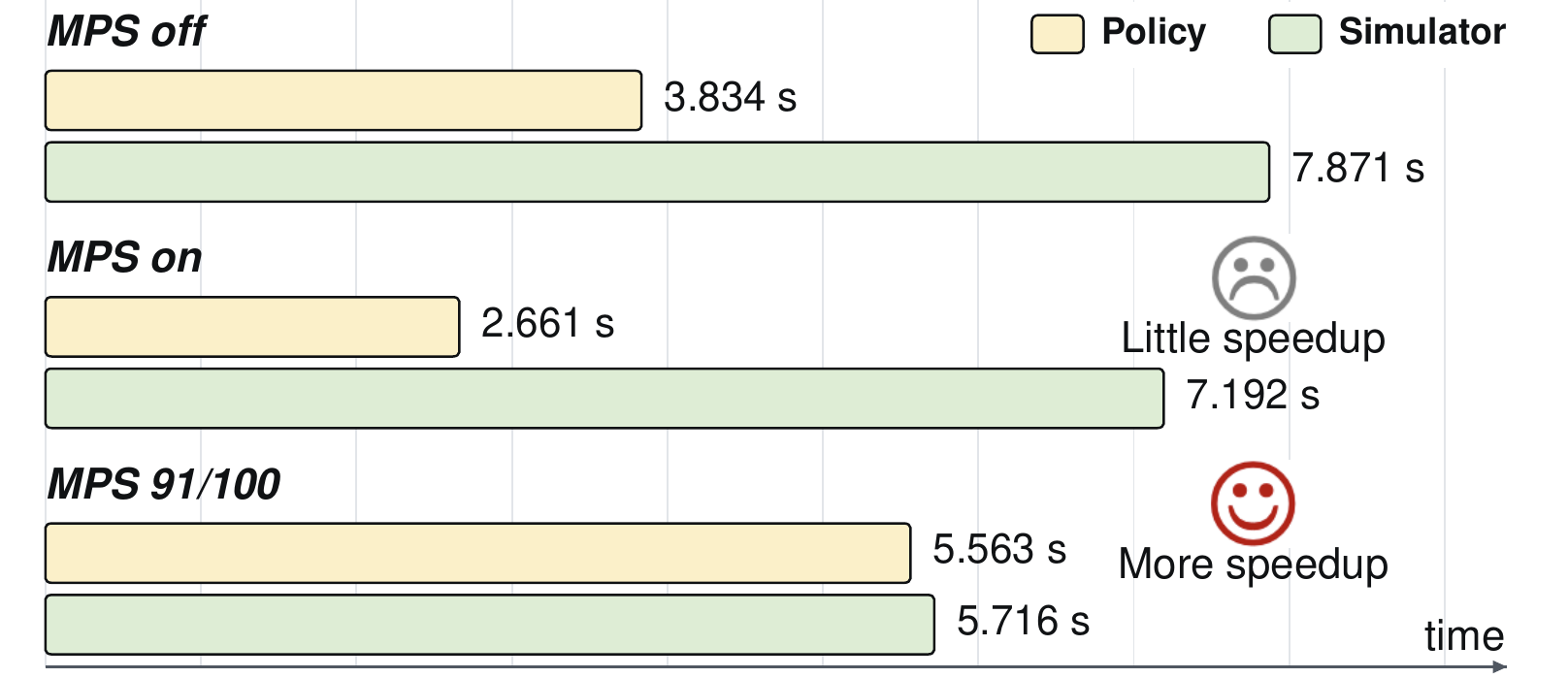}
    \vspace{-20pt}
    \caption{SM quota management and latency of overlapping execution on $\pi_{0.5}$ and ManiSkill. ``91/100'' denotes the percentage quota of policy and simulator.}
    \vspace{-15pt}
    \label{fig:SM-manage}
\end{figure}

\begin{figure*}[t]
    \centering
    \includegraphics[width=\linewidth]{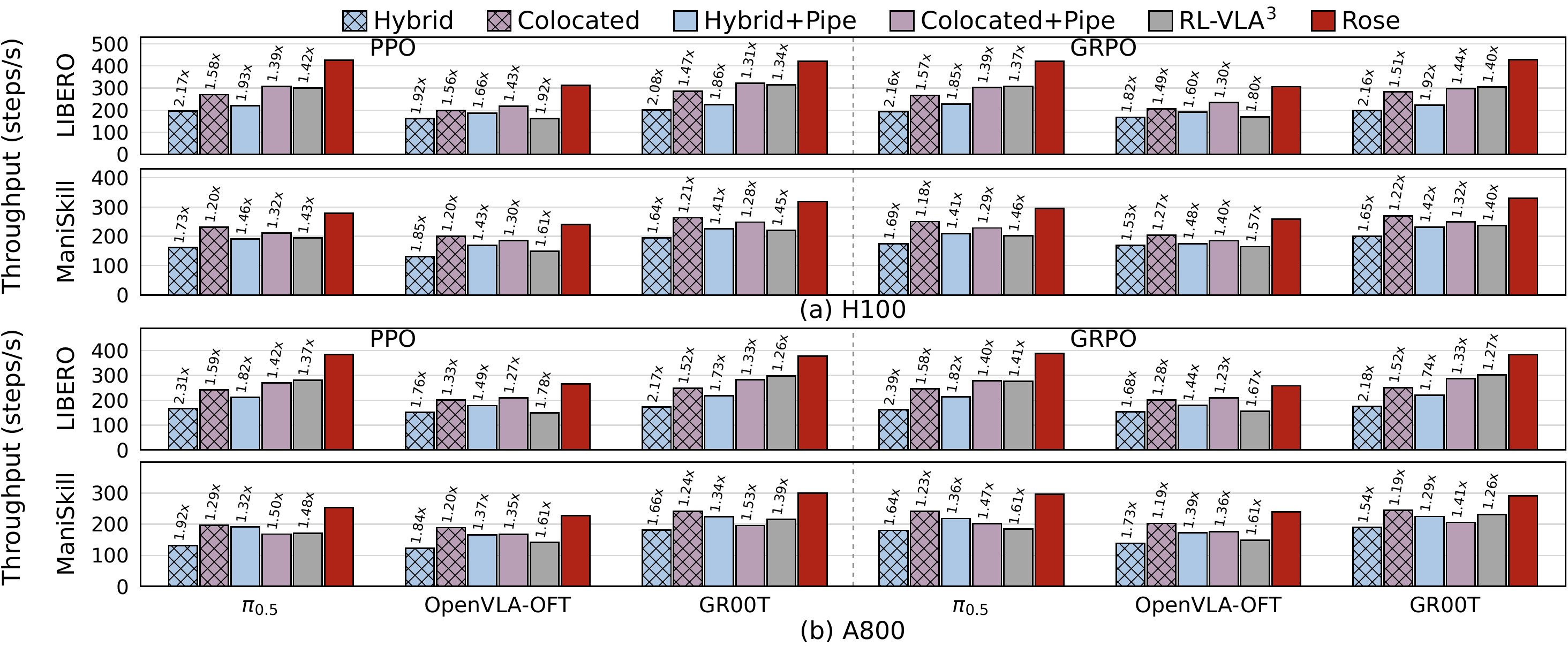}
    \vspace{-20pt}
    \caption{Training throughput.}
    \vspace{-10pt}
    \label{fig:throughput}
\end{figure*}

\subsection{Slack-aware Resource Management}
\label{sec:resource-optimization}

GPU contention can slow overlapping branches unequally while
downstream work waits for both. The earlier-ready branch can
tolerate some delay without postponing the original synchronization, and
this tolerance is its \emph{branch slack}. 
\oursys tunes the earlier-ready process's SM quota to reduce
interference, using local measurements for quota search and
complete-iteration validation for acceptance.

For readiness times $C_0 \geq F_0$ measured from a common origin,
the earlier-ready branch has slack $\sigma=C_0-F_0$.
After reducing its process quota, write the resulting readiness
times as $C_0-x$ and $F_0+y$, respectively.
The reduction in synchronization time is
$\Delta T_{\mathrm{sync}}=\min\{x,\sigma-y\}$.
Synchronization advances when the later-ready branch accelerates
($x>0$) and the earlier-ready branch's added delay remains within
its original slack ($y<\sigma$), as illustrated in Figure~\ref{fig:SM-manage}.

With $G$ fixed to the graph selected by Algorithm~\ref{alg:graph-optimization}, \oursys uses
a feasible observation $m_0$ under quotas $q_0$ to select two
overlapping branches on distinct processes sharing a GPU.
It tunes only the earlier-ready process $f$, keeping this target
and all other quotas fixed throughout the search.
If no eligible pair exists, \oursys retains $q_0$.
The comparison graph $A$ is the internal graph of the lowest
common ancestor enclosing both branches and their synchronization.

Uniform MPS provisioning caps the SM resources available to
client CUDA contexts. The backend provides $Q$ validated quota
levels (\S\ref{sec:implementation}). $q(k)$ retains $q_0$ except that $f$'s quota
is set to $k/Q$, for $k\in\{1,\ldots,Q\}$.
Algorithm~2 records in $H$ the locally measured completion time
$t(k)$ of $A$ under $(G,q(k))$, covering the same effective
work across quotas.

Algorithm~\ref{alg:resource-optimization} first reduces the quota in coarse steps of $\rho$
levels until completion time increases or the minimum quota
is reached. It then searches in both directions with unit steps
around the best coarse-search point, held fixed as the anchor;
a direction stops when its measured time exceeds the anchor's.
The measured quota with the smallest $t(k)$ is passed to
\texttt{Validate}.
Since a process quota also affects work outside $A$,
\texttt{Validate} retains the selected quota only if a complete
iteration is faster than $m_0$; otherwise, it returns $q_0$.
Measurements follow the execution semantics in \S\ref{sec:runtime-graph-optimization}.
\section{Evaluation}
\label{sec:evaluation}


We evaluate \oursys in terms of throughput, scalability, optimization effectiveness and overhead, as well as its impact on policy learning.

\subsection{Experimental Setup}
\label{sec:eval-setup}

\oursys is built on RLinf~\cite{rlinf} with 5K lines of Python code, adding
\texttt{ExecutionNode} and \texttt{ExecutionTrace} wrappers, and
graph transformation implementation.
It applies GPU quotas through MPS provisioning at a granularity of two SMs, as detailed in \S\ref{sec:implementation}. 


\begin{table}[t]
    \centering
    \vspace{-5pt}
    \caption{GPU execution efficiency on A800.}
    \label{tab:gpu-efficiency}
    \small
    \begin{tabular}{llcc}
        \toprule
        \multirow{2}{*}{Model}
        & \multirow{2}{*}{Simulator}
        & \multicolumn{2}{c}{Average SM activity} \\
        \cmidrule(lr){3-4}
        & & Baseline & \oursys \\
        \midrule
        $\pi_{0.5}$ & LIBERO
        & 42.45\% & 59.15\%
        \\
        & ManiSkill
        & 46.38\% & 58.01\%
        \\
        \midrule
        OpenVLA-OFT & LIBERO
        & 54.98\% & 68.08\%
        \\
        & ManiSkill
        & 43.59\% & 56.57\%
        \\
        \bottomrule
    \end{tabular}
\end{table}

\noindent \textbf{Testbeds.}
We evaluate throughput on two 8-GPU platforms: NVIDIA H100 GPUs with Intel Xeon Platinum 8481C CPUs, and NVIDIA A800 GPUs with Intel Xeon Platinum 8358P CPUs. Each GPU has 80\,GB memory, and each node provides 920\,GB host memory. Scaling experiments use 4 and 8 A800 GPUs within one node and 16 GPUs across two nodes, connected via NVLink within nodes and 8$\times$200\,Gbps RDMA NICs across nodes.


\noindent \textbf{Workloads and metrics.} We evaluate $\pi_{0.5}$~\cite{pi05}, OpenVLA-OFT~\cite{openvla_oft}, and GR00T N1.6~\cite{groot} on LIBERO-Spatial~\cite{libero} and ManiSkill~\cite{maniskill3} (``PutOnPlateInScene25Main-v3''), using PPO~\cite{ppo} and GRPO~\cite{deepseekmath}. Across systems, we match rollout volumes and policy-update budgets using benchmark-recommended configurations. We report end-to-end throughput (simulator steps/s) and average SM activity across GPUs over each iteration.

\noindent \textbf{Baselines.} We compare \oursys against four RLinf configurations and RL-VLA$^3$~\cite{RL-VLA3}.
The RLinf configurations are explained in section~\ref{sec:background} and denoted as
\textit{Colocated},
\textit{Colocated+Pipe},
\textit{Hybrid}, and
\textit{Hybrid+Pipe}.

\subsection{End-to-End Performance}
\label{sec:eval-e2e}

\noindent \textbf{Training throughput.} Figure~\ref{fig:throughput} reports throughput for the twelve workload configurations on H100
and A800.

(1) H100. Across PPO and GRPO, \oursys achieves up to
1.39$\times$,
1.43$\times$, and
1.44$\times$
the RLinf throughput with the best configuration on LIBERO for $\pi_{0.5}$,
OpenVLA-OFT, and GR00T N1.6, respectively.
On ManiSkill, the corresponding speedups
are 1.20$\times$,
1.27$\times$, and
1.22$\times$.
(2) A800. Across PPO and GRPO, \oursys achieves up to
1.42$\times$,
1.27$\times$, and
1.33$\times$
the RLinf throughput on LIBERO for the same models.
On ManiSkill, the corresponding maximum speedups
are 1.29$\times$,
1.20$\times$, and
1.24$\times$.
(3) \oursys also achieves up to 1.92$\times$ and 1.61$\times$ the throughput of RL-VLA$^3$ on LIBERO and ManiSkill across platforms respectively.

We observe that enabling simulator-group pipelining reduces throughput in RLinf's colocated configuration
on ManiSkill.
Grouping environments reduces the batch size processed by each component.
Policy inference time decreases approximately linearly with batch size, whereas the less compute-intensive GPU simulation sees only a modest reduction.
Executing the additional simulator group therefore
substantially increases cumulative simulation time,
making simulation the bottleneck.
As discussed in Section~\ref{sec:motivation}, this configuration still time-slices GPU execution across policy and simulator processes, thus providing insufficient
concurrent GPU execution to offset the increased
simulation cost, resulting in lower throughput.

\begin{figure}[t]
    \centering
    \includegraphics[width=\linewidth]{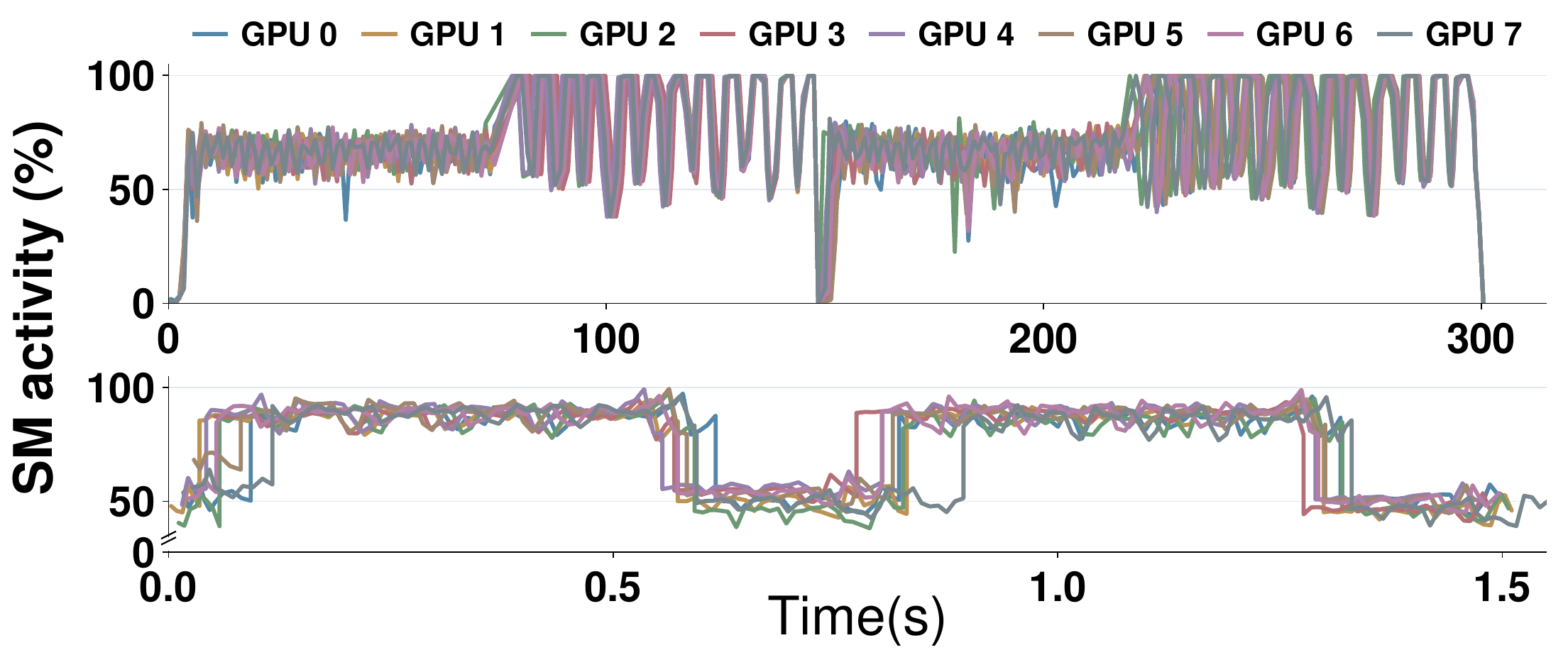}
    \vspace{-20pt}
    \caption{SM activity after \oursys optimization over a training iteration. The lower panel is spread from part of the upper rollout.}
    \vspace{-10pt}
    \label{fig:SM-final}
\end{figure}

\noindent \textbf{GPU execution efficiency.} We plot the SM activity for a complete training iteration after automatic optimization in figure~\ref{fig:SM-final}. 
SM activity remains below 50\% for 8.4\% of the interaction-loop execution time and 21.3\% of the complete iteration time, corresponding to relative reductions of 78.4\% and 64.3\% in these time fractions, respectively.
The remaining low-activity windows persist because no complementary and dispatchable work is available to fill them.
Table~\ref{tab:gpu-efficiency} compares average SM activity between \oursys and RLinf with the best configuration for $\pi_{0.5}$ and OpenVLA-OFT on LIBERO and ManiSkill.
Across configurations, \oursys improves SM activity by 23.8--39.3\% on average.

\begin{figure}[t]
    \centering
    \includegraphics[width=\linewidth]{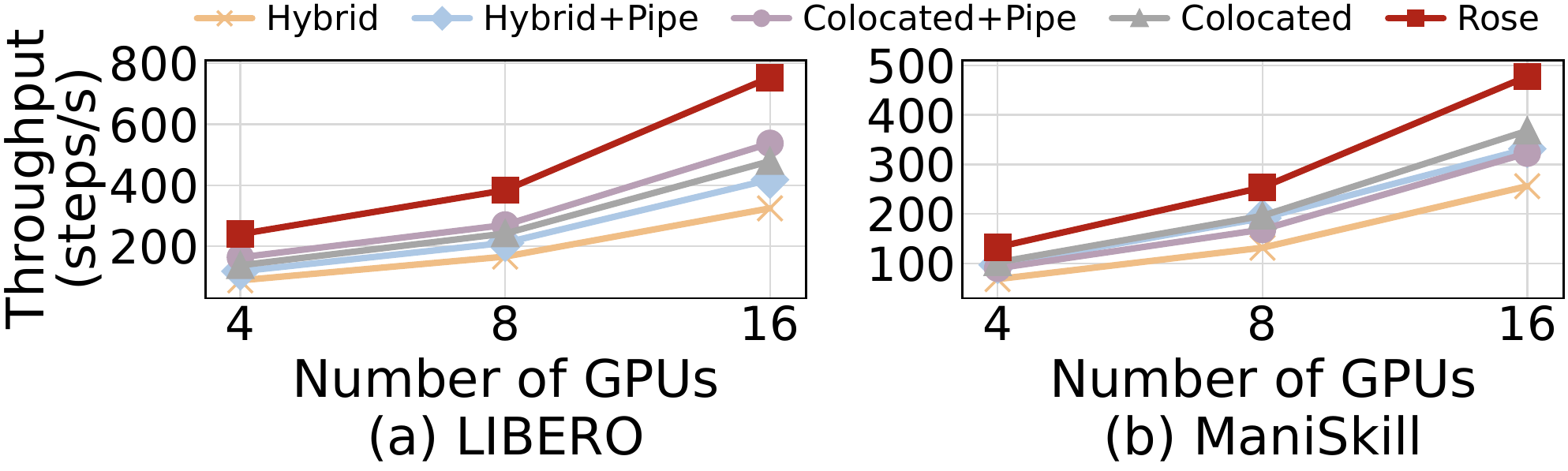}
    \vspace{-20pt}
    \caption{Training throughput on 4-16 GPUs.}
    \vspace{-10pt}
    \label{fig:scaling}
\end{figure}

\noindent \textbf{Scaling.} We evaluate scaling on four, eight, and sixteen A800 GPUs on $\pi_{0.5}$ PPO training with two simulators.
The workload for each rank remains fixed, so the total simulator and training workloads increase with the GPU count under data parallelism.
Figure~\ref{fig:scaling} compares \oursys with different RLinf configurations for each
simulator and GPU count.
On LIBERO, \oursys improves throughput over the best configuration by 46.5\%, 42.2\%, and 40.2\% on four, eight, and sixteen GPUs, respectively.
On ManiSkill, the corresponding improvements are 31.5\%, 29.2\%, and 29.8\%.

\subsection{Ablation Study}
\label{sec:eval-ablation}

\begin{figure}[t]
    \centering
    \includegraphics[width=\linewidth]{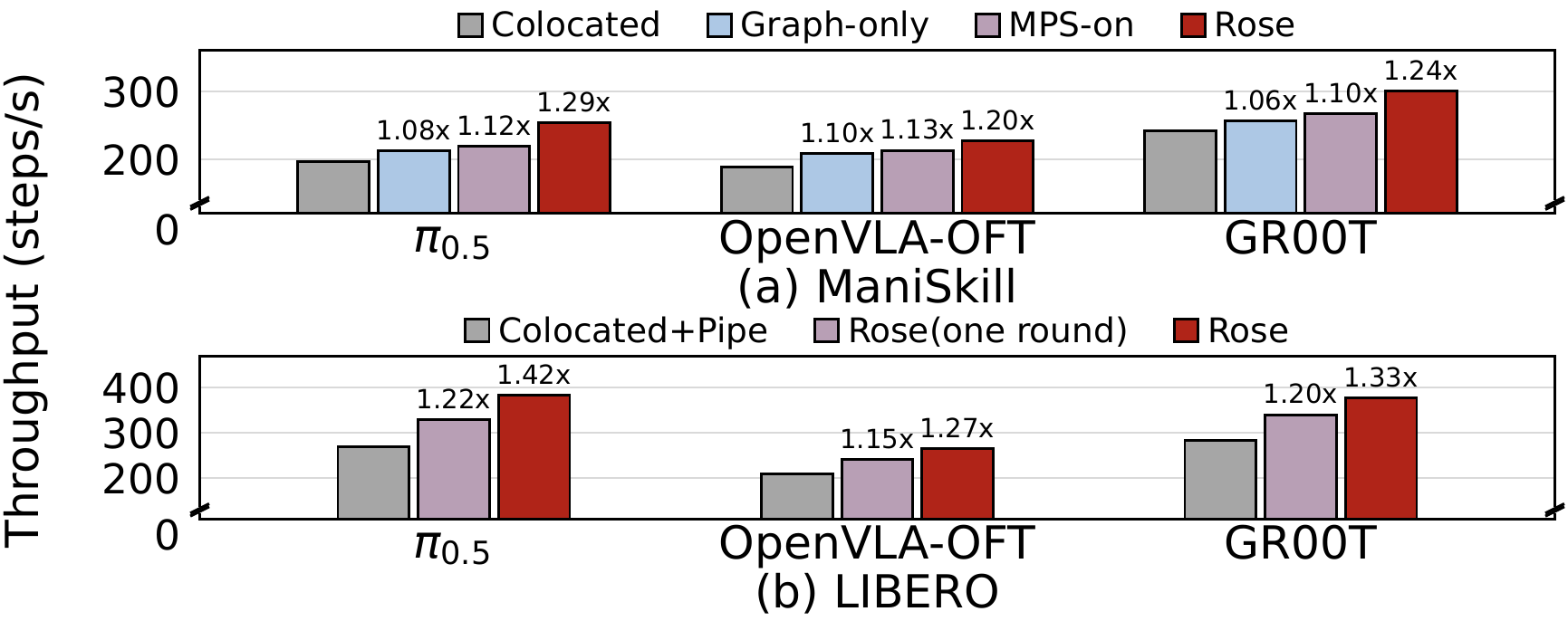}
    \vspace{-20pt}
    \caption{Training throughput ablation.}
    \vspace{-10pt}
    \label{fig:ablation}
\end{figure}

\begin{figure}[t]
    \centering
    \includegraphics[width=\linewidth]{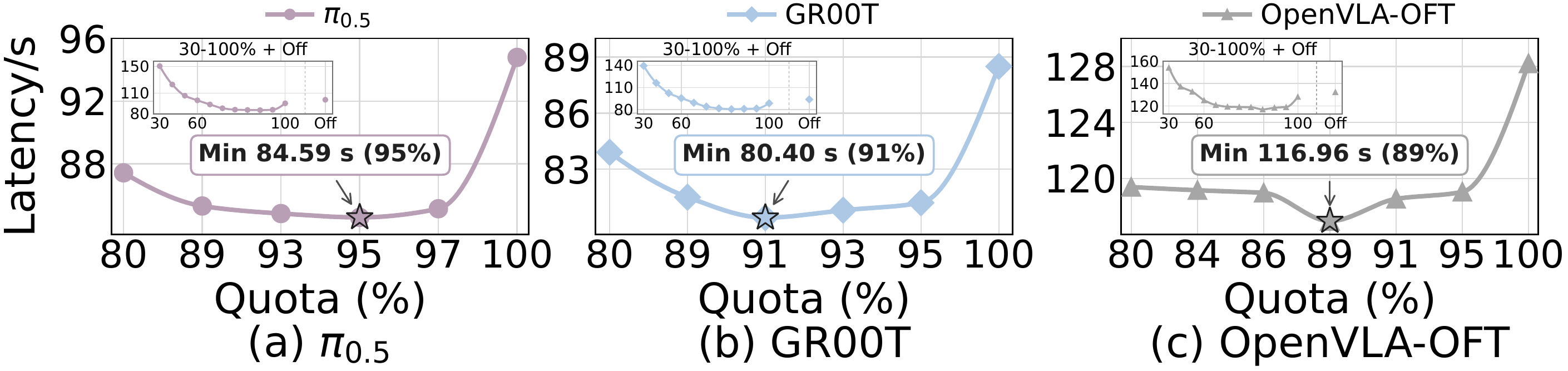}
    \vspace{-15pt}
    \caption{Latency during quota tuning.}
    \vspace{-10pt}
    \label{fig:quota}
\end{figure}

We evaluate MPS and quota optimization on ManiSkill using the same workload and component placement:
(1) \textbf{Colocated}, the best RLinf configuration;
(2) \textbf{Graph-only}, with graph optimization but no MPS;
(3) \textbf{MPS-on}, with the optimized graph and default MPS quotas; and
(4) \textbf{Full \oursys}, which additionally tunes the selected process's SM quota.

Figure~\ref{fig:ablation} (a) reports the throughput
of the four variants.
Graph optimization alone achieves
1.08$\times$, 1.10$\times$ and 1.06$\times$ the throughput of Colocated.
With MPS enabled, the speedup becomes 1.12$\times$, 1.13$\times$ and 1.10$\times$, increasing by 3.4\%, 2.3\% and 3.9\%.
Finally, SM quota optimization brings the overall speedup of Full \oursys to
1.29$\times$, 1.20$\times$ and 1.24$\times$, increasing by 15.6\%, 6.7\% and 12.6\%.
All speedups use Colocated as the baseline.
The speedup breakdown of three methods is 33.1\%, 13.5\%, 53.4\% for $\pi_{0.5}$, 54.9\%, 12.1\%, 33.0\% for OpenVLA-OFT, and 29.2\%, 18.3\%, 52.5\% for GR00T.

Figure~\ref{fig:quota} shows how local latency varies while tuning SM quota for all models.
Local latency is measured over an overlap execution window spanning four action chunks.
The x-axis denotes the SM quota of the selected process with slack.
The full-range sweep shows that overly restrictive quotas substantially increase latency, while the default quota (100\%) is suboptimal. Zooming into the 80–100\% range reveals a U-shaped latency curve, with minima at 95\%, 91\%, and 89\%, respectively.

We additionally evaluate the benefit of using multiple graph optimization rounds. There are three variants:
(1) \textbf{Colocated+Pipe.}
The best configuration for RLinf on LIBERO.
(2) \textbf{\oursys (one round).}
Execution graph after optimization for one round.
(3) \textbf{Full \oursys.}
Execution graph after completing two rounds of optimization.
Figure~\ref{fig:ablation} (b) reports the throughput of the variants.
The first and second rounds contribute 60.7\% and 39.3\% of the total reduction in iteration time for $\pi_{0.5}$, 62.9\% and 37.1\% for OpenVLA-OFT, and 67.6\% and 32.4\% for GR00T, respectively.

\subsection{Optimization Overhead}
\label{sec:eval-overhead}

We measure optimization overhead with the target limit set to $K=3$.
For both workloads, tuning includes an end-to-end reference profile and two rounds of graph optimization, each followed by end-to-end profiling.
ManiSkill additionally requires seven short MPS quota trials and a final validation.
We distinguish total optimization time from net overhead by subtracting training execution whose progress is retained.
For $\pi_{0.5}$ with PPO on LIBERO, two rounds of graph optimization take 1,998\,s, including 1,336\,s of training execution, leaving 662\,s of net overhead.
The policy reaches a 90.2\% success rate after 252 iterations in 85,727\,s, including optimization.
Total time and net overhead therefore account for 2.3\% and 0.8\% of the complete run, respectively.
For ManiSkill, the total optimization time is 4,615\,s, including 2,290\,s of retained training execution and 2,325\,s of net overhead.
Relative to its complete training time of 100,716\,s to reach a 90.0\% success rate, total time and net overhead correspond to 4.6\% and 2.3\%, respectively.

\section{Conclusion}

We presented \oursys, a runtime-aware system for automatic overlap optimization in embodied RL.
Combining a programmable interface with runtime measurements, \oursys performs two-stage optimization through a window-guided graph optimizer and a slack-aware resource manager while preserving training semantics.
Evaluations demonstrate substantial improvements in training throughput and GPU utilization.

\bibliographystyle{ACM-Reference-Format}
\bibliography{reference}

\clearpage
\appendix
\section{Runtime}
\noindent \textbf{Runtime characteristics.}
The data plane records host-call duration for every invocation
and additionally records device time when available.
Separately, it maintains a device-wide SM activity
(\cite{nvidia_dcgm_profiling}) trace throughout the iteration
for the GPU associated with each rank.
For an invocation with device activity, the duration extends
from the host start to the later of host return and device completion.
Parent and child nodes provide alternative observation scopes
but their measurements are not added as independent work.

\noindent \textbf{Replacement of callable graph.}
The runtime implements a node’s transformed internal graph as an alternative callable.
A transformed graph must execute the node’s complete effective work while preserving its output interface and state effects.
A hook at the node’s invocation entry selects this callable when registered.
Each invocation executes either the native graph or its replacement, never both.

\noindent \textbf{Dependency-driven execution.}
An invocation is submitted only when its input, state,
and scheduling requirements are satisfied.
Output readiness accounts for required device computation and communication.
A host return alone does not establish GPU output readiness.
Intermediate buffers remain live until their required consumers complete.

\noindent \textbf{Local validation.}
Local validation compares the reference and transformed
graphs over the same workload with comparable
input and mutable state prepared.
Preparation is excluded from regional timing but included in optimization overhead.
Local trials do not commit their mutations to the ongoing training state.

\section{Implementation}
\label{sec:implementation}

\oursys extends RLinf~\cite{rlinf} with 5K lines of Python, adding
execution-node wrappers, the \texttt{ExecutionTrace} data plane, graph transformations
and runtime support.

\noindent \textbf{Execution integration and bindings.}
\oursys adds an invocation hook to the common
\texttt{ExecutionNode} wrapper for synchronous and asynchronous execution.
At each invocation, the hook selects the alternative callable registered for the node or its native
entry if no alternative is registered.
The driver installs these callables and their scheduling
metadata on participating workers and launches invocations
through RLinf's distributed execution interface.

Each replacement callable returns results through the original output interface.
\emph{Spatial split} binds invocations to subsets of the
original inputs.
\emph{Temporal split} preserves data and state bindings for separate graphs.
Fragment outputs retain their positions in the complete
result, which Merge uses to reconstruct downstream inputs.

\noindent \textbf{Execution scheduling and completion.}
Independent execution graphs run as separate asynchronous
tasks within each worker.
Before execution, tasks wait for the required input, state and scheduling dependencies while allowing other runnable tasks to progress.
Cross-worker data transfers reuse RLinf's routed
communication.

For asynchronous entries, the wrapper invokes \texttt{trace.after} from the returned future's completion
callback.
Output readiness is tracked separately through backend
completion events covering device work and required
data transfers.

\noindent \textbf{MPS provisioning and local quota trials.}
\oursys applies quotas through uniform MPS provisioning within each GPU worker process~\cite{nvidia_mps}.
This ensures quota to take effect for simulator backend that manages CUDA contexts internally such as PhysX, which a Python-level Green Context wrapper cannot reliably constrain.
Before launching a worker, \oursys sets \texttt{CUDA\_\allowbreak MPS\_\allowbreak ACTIVE\_\allowbreak THREAD\_\allowbreak PERCENTAGE}.
The resulting limit remains fixed for the client's lifetime, so quota changes require replacing the affected
test worker.
Quota trials use short workloads on separately initialized
workers.
We characterize effective quota levels by querying
\texttt{cudaDevAttr\allowbreak MultiProcessorCount} in fresh MPS
clients under different percentage settings.
On the tested Hopper and Ampere GPUs, the smallest observed SM-count increment was two.

\noindent \textbf{Profiling and timing.}
Local validation measures execution time without Nsight Systems profiling.
Within each comparison, reference and candidate runs
use the same workload, timing boundaries, and profiling
mode.

During profiled iterations, Nsight Systems~\cite{nvidia_nsight} collects
device-wide SM activity throughout the iteration.
NVTX ranges mark exposed invocations only in the first
and last few steps of inner loops to reduce recording
overhead.
Invocation identifiers map these ranges to structural
nodes, allowing sampled execution intervals to be aligned
with the SM-activity trace.
For unsampled invocations, device-associated durations
are estimated from sampled invocations of the same node.

Completion time of the comparison graph is the wall-clock span from its designated start to its completion event, including computation, data transfer and synchronization.
End-to-end timing uses complete-iteration boundaries.

\noindent \textbf{Optimization overhead.}
Optimization overhead includes worker and model initialization, graph transformation and installation, data and state preparation, local trials and complete iteration validation.
Quota tuning additionally incurs worker replacement
and reinitialization, while compatible graph trials
reuse initialized workers.
These costs are reported in Section~\ref{sec:evaluation}.

\end{document}